\documentstyle[12pt]{article}
\begin{document}
\thispagestyle{empty}
\newcommand{\be}{\begin{equation}}
\newcommand{\ee}{\end{equation}}
\newcommand{\sect}[1]{\setcounter{equation}{0}\section{#1}}
\newcommand{\vs}[1]{\rule[- #1 mm]{0mm}{#1 mm}}
\newcommand{\hs}[1]{\hspace{#1mm}}
\newcommand{\mb}[1]{\hs{5}\mbox{#1}\hs{5}}
\newcommand{\bea}{\begin{eqnarray}}
\newcommand{\eea}{\end{eqnarray}}
\newcommand{\wt}[1]{\widetilde{#1}}
\newcommand{\ux}[1]{\underline{#1}}
\newcommand{\ov}[1]{\overline{#1}}
\newcommand{\sm}[2]{\frac{\mbox{\footnotesize #1}\vs{-2}}
           {\vs{-2}\mbox{\footnotesize #2}}}
\newcommand{\prt}{\partial}
\newcommand{\eps}{\epsilon}\newcommand{\p}[1]{(\ref{#1})}
\newcommand{\R}{\mbox{\rule{0.2mm}{2.8mm}\hspace{-1.5mm} R}}
\newcommand{\Z}{Z\hspace{-2mm}Z}
\newcommand{\cd}{{\cal D}}
\newcommand{\cg}{{\cal G}}
\newcommand{\ck}{{\cal K}}
\newcommand{\cw}{{\cal W}}
\newcommand{\vj}{\vec{J}}
\newcommand{\vl}{\vec{\lambda}}
\newcommand{\vz}{\vec{\sigma}}
\newcommand{\vt}{\vec{\tau}}
\newcommand{\poiss}{\stackrel{\otimes}{,}}
\newcommand{\tx}{\theta_{12}}
\newcommand{\tb}{\overline{\theta}_{12}}
\newcommand{\zw}{{1\over z_{12}}}
\newcommand{\sqp}{{(1 + i\sqrt{3})\over 2}}
\newcommand{\sqm}{{(1 - i\sqrt{3})\over 2}}
% REVUES POUR BIBLIO
\newcommand{\NP}[1]{Nucl.\ Phys.\ {\bf #1}}
\newcommand{\PLB}[1]{Phys.\ Lett.\ {B \bf #1}}
\newcommand{\PLA}[1]{Phys.\ Lett.\ {A \bf #1}}
\newcommand{\NC}[1]{Nuovo Cimento {\bf #1}}
\newcommand{\CMP}[1]{Commun.\ Math.\ Phys.\ {\bf #1}}
\newcommand{\PR}[1]{Phys.\ Rev.\ {\bf #1}}
\newcommand{\PRL}[1]{Phys.\ Rev.\ Lett.\ {\bf #1}}
\newcommand{\MPL}[1]{Mod.\ Phys.\ Lett.\ {\bf #1}}
\newcommand{\BLMS}[1]{Bull.\ London Math.\ Soc.\ {\bf #1}}
\newcommand{\IJMP}[1]{Int.\ J.\ Mod.\ Phys.\ {\bf #1}}
\newcommand{\JMP}[1]{Jour.\ Math.\ Phys.\ {\bf #1}}
\newcommand{\LMP}[1]{Lett.\ Math.\ Phys.\ {\bf #1}}
%\renewcommand{\thefootnote}{\fnsymbol{footnote}}
%\footnotemark
\newpage
\setcounter{page}{0} \pagestyle{empty} \vs{12}
%{\large{\em Draft version}}
\begin{center}
{\LARGE {\bf An $N=8$ Superaffine Malcev Algebra}}\\ {\quad}\\
{\LARGE{\bf and Its $N=8$ Sugawara}}\\
 [0.8cm]

\vs{10} {\large H.L. Carrion, M. Rojas and F. Toppan} ~\\ \quad
\\
 {\large{\em CBPF - CCP}}\\{\em Rua Dr. Xavier Sigaud
150, cep 22290-180 Rio de Janeiro (RJ)}\\{\em Brazil}\\

\end{center}
{\quad}\\
\centerline{ {\bf Abstract}}

\vs{6}

A supersymmetric affinization of the algebra of octonions is
introduced. It satisfies a super-Malcev property and is $N=8$
supersymmetric. Its Sugawara construction recovers, in a special
limit, the non-associative $N=8$ superalgebra of Englert et
al.\par This paper extends to supersymmetry the results obtained
by Osipov in the bosonic case.

\vs{6} \vfill \rightline{CBPF-NF-011/01} {\em E-mails:}{
lenyj@cbpf.br, mrojas@cbpf.br, toppan@cbpf.br}
%\newpage
\pagestyle{plain}
\renewcommand{\thefootnote}{\arabic{footnote}}
%\setcounter{footnote}{0}
%\vs{8}

\section{Introduction.}

It is well-known that a connection exists (see e.g. \cite{KuTo}),
between the division algebras of complex numbers, quaternions,
octonions  and the $N$-extended supersymmetries, for values
$N=2,4,8$ respectively. However, the characterization of
$N$-extended supersymmetries in terms of division algebras is not
always worked out explicitly. This is especially true in the case
of the octonions since, being both non-commutative and
non-associative, they present peculiar difficulties.\par In the
present work we extend to the supersymmetric case the approach and
results obtained by Osipov some years ago in a series of three
papers concerning the affinization  \cite{{Osi1},{Osi2}} of the
octonionic algebra and its Sugawara \cite{Osi1} construction.\par
We postpone to the Conclusions (where a list of possible
applications of the results here obtained is furnished) the
discussion about the motivations of our paper. In this
Introduction we limit ourselves to summarize the main results of
the present work.\par The algebra of octonions is {\em
supersymmetrically affinized} following a proposal made in
\cite{{Osi1},{Osi2}}. While in \cite{Osi1} the bosonic case only
was considered, in \cite{Osi2} a superaffinization was proposed.
The superaffinization introduced here however, formulas
(\ref{eleven}) and (\ref{fourteen}), differs from the one in
\cite{Osi2}. Our formulation is manifestly supersymmetric, while
the one of \cite{Osi2} is not.\par Later we explicitly prove that
the superaffine octonionic algebra (\ref{eleven}) is superMalcev,
i.e. it satisfies a graded version of the Malcev identity.\par
Despite being expressed in terms of manifestly $N=1$ superfields,
the (\ref{eleven}) superaffine algebra is $N=8$ supersymmetric.
One way of seeing this is in consequence of the existence of a
supersymmetric Sugawara construction given in formulas
(\ref{sixteen}). This Sugawara realization is a generalization of
both the Osipov's construction \cite{Osi1} in the purely bosonic
case (all fermionic fields set equal to zero), as well as the
$N=4$ S.C.A. Sugawara of reference \cite{IKT}, recovered when just
the (superaffinized version of the) quaternionic subalgebra of the
octonions is taken into account.
\par
Unlike the purely bosonic \cite{Osi1} and the $N=4$ S.C.A.
\cite{IKT} Sugawaras, the Sugawara-induced fields (\ref{sixteen})
do not close a superalgebra due to the presence of extra-terms
(dependent on fermionic fields associated to octonionic
structure-constants not belonging to a quaternionic subalgebra).
However, after a suitable limiting procedure is taken into
account, a closed $N=8$ generalization of the Virasoro algebra
(with one bosonic spin-$2$ field, $8$ fermionic spin-$\frac{3}{2}$
and $7$ bosonic spin-$1$ fields) is recovered. It corresponds to
the so-called ``Non-associative $N=8$ S.C.A." introduced for the
first time by Englert et al. in reference \cite{ESTvPS}.
Therefore, as a byproduct of our investigation concerning the
superaffinization of the algebra of the octonions and its Sugawara
realization, we found as a bonus its connection with a very
remarkable and quite ``mysterious" superconformal algebra
appearing in the literature. The latter algebra is not of
(super)Malcev type. This point will be commented in the text.\par
In order to obtain our results we made extensive use of the
Thielemans' package for classical OPE computations with
Mathematica. We developed our own special package to deal with
octonions.

\vspace{0.2cm}\noindent{\section{Notations and preliminary
results}}

In this section the basic properties of the division algebra of
the octonions, following \cite{GuKe}, are reviewed. They are later
used for our construction. \par A generic octonion $x$ is
expressed as $x=x_a\tau_a$ (throughout the text the convention
over repeated indices, unless explicitly mentioned, is
understood), where $x_a$ are real numbers while $\tau_a$ denote
the basic octonions, with $a=0,1,2,...,7$.\par $\tau_0\equiv {\bf
1}$ is the identity, while $\tau_\alpha$, for $\alpha =1,2,...,7$,
denote the imaginary octonions. In the following a Greek index is
employed for imaginary octonions, a Latin index for the whole set
of octonions (identity included).\par The octonionic
multiplication can be introduced through
\begin{eqnarray}
\tau_\alpha \cdot \tau_\beta &=& -\delta_{\alpha\beta} \tau_0 +
C_{\alpha\beta\gamma} \tau_\gamma , \label{one}
\end{eqnarray}
with $C_{\alpha\beta\gamma}$ a set of totally antisymmetric
structure constants which, without loss of generality, can be
taken to be
\begin{eqnarray}
&C_{123}=C_{147}=C_{165}=C_{246}=C_{257}= C_{354}=C_{367}=1.&
\label{two}
\end{eqnarray}
and vanishing otherwise.\par It is also convenient to introduce,
in the seven-dimensional imaginary octonions space, a $4$-indices
totally antisymmetric tensor $C_{\alpha\beta\gamma\delta}$, dual
to $C_{\alpha\beta\gamma}$, through \begin{eqnarray}
C_{\alpha\beta\gamma\delta} &=&
\frac{1}{6}\varepsilon_{\alpha\beta\gamma\delta\epsilon\zeta\eta}
C_{\epsilon\zeta\eta} \label{three}
\end{eqnarray}
(the totally antisymmetric tensor
$\varepsilon_{\alpha\beta\gamma\delta\epsilon\zeta\eta}$ is
normalized so that $\varepsilon_{1234567}=+1$). \par The
octonionic multiplication is not associative since for generic
$a,b,c$ we get$(\tau_a\cdot \tau_b)\cdot \tau_c \neq \tau_a \cdot
(\tau_b\cdot\tau_c)$. However the weaker condition of
alternativity is satisfied. This means that, for $a=b$, the
associator
\begin{eqnarray}
\relax [ \tau_a,\tau_b,\tau_c]\equiv (\tau_a\cdot\tau_b)\cdot
\tau_c -\tau_a\cdot(\tau_b\cdot \tau_c) \label{four}
\end{eqnarray}
is vanishing.\par We further introduce the commutator algebra of
octonions through
\begin{eqnarray}
\relax [\tau_a,\tau_b] &=_{def}
\tau_a\cdot\tau_b-\tau_b\cdot\tau_a = f_{abc}\tau_c, \label{five}
\end{eqnarray}
$f_{abc}$ are the structure constants and can be read from
(\ref{one}); we have $f_{abc} = 2 C_{abc}$ (where $C_{abc}$
coincides with $C_{\alpha\beta\gamma}$ for $a,b,c=1,2,...,7$ and
is vanishing otherwise).\par The above-defined commutator brackets
$[.,.]$ satisfy the two properties below, which make the
commutator algebra a Malcev-type algebra
\begin{eqnarray}
\relax [{\bf x},{\bf x}] &=& 0,\nonumber\\ \relax J({\bf x},{\bf
y}, [{\bf x},{\bf z}]) &=& [J({\bf x},{\bf y},{\bf z}),{\bf x}]
\label{six}
\end{eqnarray}
for any given triple ${\bf x},{\bf y},{\bf z}$ of octonions. \par
$J({\bf x},{\bf y},{\bf z})$ is the Jacobian
\begin{eqnarray}
\relax J({\bf x},{\bf y},{\bf z}) &=& [[{\bf x},{\bf y}],{\bf z}]
]+[[{\bf y},{\bf z}],{\bf x}] + [[{\bf z},{\bf x}],{\bf y}].
\label{seven}
\end{eqnarray}
The second relation in (\ref{six}) is in consequence of the
alternativity property.\par Malcev algebras are a special
generalization of the Lie algebras, obtained by relaxing the
condition of the vanishing of the Jacobian.

\vspace{0.2cm}\noindent{\section{The supersymmetric affinization
of the octonionic algebra.}}

In this section we introduce the superaffinization of the
commutator algebra of the octonions (\ref{five}) or, in short, the
superaffine octonionic algebra, which for later convenience will
be denoted as ${\widehat{\cal O}}$-algebra. Since the commutator
algebra of octonions (\ref{five}) is not a Lie algebra, its
superaffinization implies generalizing the concept of
superaffinization of a given Lie algebra ${\cal G}$. However, the
standard notion of superaffinization should be recovered when
formulas are specialized to the subalgebra of quaternions, which
is equivalent to the Lie algebra $sl(2)\oplus o(2)$.\par Given a
Lie algebra ${\cal G}$ with generators $g_i$ and structure
constants $f_{ijk}$, the superaffine Lie algebra ${\widehat{\cal
G}}$ is introduced by associating a bosonic spin-$1$ field
$j_i(x)$ and a spin-$\frac{1}{2}$ fermionic field $\psi_i(x)$ to
each Lie algebra generator $g_i$. These fields can be thought as
components of a single fermionic $N=1$ superfield $\Psi_i(X)
=\psi_i(x)+\theta j_i(x)$, where $X\equiv (x,\theta)$ denotes a
superspace coordinate and $\theta$ is a Grassmann variable
satisfying $\theta^2=0$. The superaffine ${\widehat{\cal G}}$
algebra is introduced through the brackets
\begin{eqnarray}
\{\Psi_i(X),\Psi_j(Y)\} &=& f_{ijk} \Psi_k(Y)\delta(X,Y) + k\cdot
tr(g_ig_j)D_Y\delta (X,Y). \label{eight}
\end{eqnarray}
In the formula above $X\equiv x,\theta$ and $Y\equiv y, \eta$ are
superspace coordinates. $\delta(X,Y)$ is the supersymmetric
delta-function
\begin{eqnarray} \delta(X,Y)
&=&\delta(x-y)(\theta-\eta) \label{nine} \end{eqnarray} and $D_Y$
the supersymmetric derivative
\begin{eqnarray}
D_Y &=& \frac{\partial}{\partial\eta} +\eta\partial_y. \label{ten}
\end{eqnarray}
The last term in the r.h.s. of (\ref{eight}) corresponds to the
central extension (in its absence we obtain the superloop
algebra). The trace is taken in a given representation of ${\cal
G}$ (let's say the adjoint). The formula (\ref{eight}) is
manifestly supersymmetric since it is constructed with $N=1$
superfields and covariant quantities. \par In the absence of the
central extension, the construction of the superloop extension of
the (\ref{five}) octonionic algebra is straightforward. The
construction of the central extension is however delicate. In this
case it is no longer possible to introduce it by making reference
to a matrix representation since no such representation exists for
the non-associative algebra of the octonions. Our proposal
consists in introducing the superaffine algebra ${\widehat{\cal
O}}$, based on the fermionic superfields $\Psi_a(X)$,
$a=0,1,2,...,7$, through the position
\begin{eqnarray}
\{\Psi_a(X),\Psi_b(Y)\} &=& f_{abc} \Psi_c(Y)\delta(X,Y) + k\cdot
\Pi(\tau_a\cdot\tau_b)D_Y\delta (X,Y), \label{eleven}
\end{eqnarray}
where $\Pi(\tau_a\cdot\tau_b)$ denotes the projection over the
identity ${\bf 1}$ in the composition law.\par Some comments are
in order. With the above definition the correct (anti-)symmetry
properties of the brackets are satisfied. The algebra is
manifestly $N=1$ supersymmetric and, when the formulas are
specialized to the quaternionic subalgebra, we recover the
standard superaffinization commented above. Moreover the Osipov's
bosonic construction is recovered when setting the fermionic
fields equal to zero.
\par We therefore regard (\ref{eleven}) as the correct
superaffinization of the commutator algebra of the octonions. It
should be noticed that a superaffinization has been proposed in
\cite{Osi2} in terms of component fields. However the algebra
introduced in \cite{Osi2} presents a central extension which fails
to be manifestly supersymmetric, in contrast with the
(\ref{eleven}) algebra.\par The superaffine ${\widehat{\cal O}}$
algebra is superMalcev, as a simple inspection can prove, i.e. it
satisfies a ${\bf Z}_2$-graded extension of the Malcev properties,
where the brackets are now ${\bf Z}_2$-graded and the Jacobian is
replaced by a superJacobian. The $\epsilon_{\bf x}$ grading of the
${\bf x}$ field is either $0$ or $1$ according to the bosonic
(respectively fermionic) character of ${\bf x}$. The graded
brackets are defined as
\begin{eqnarray}
\relax [{\bf x},{\bf y}] &=&(-1)^{\epsilon_{\bf x}\epsilon_{\bf y}
+1} [{\bf y},{\bf x}] \label{twelve}
\end{eqnarray}
(${\bf x},{\bf y}$ denote graded-algebra elements like fields or
superfields; in the following the graded-brackets are denoted
either as ``[.,.]" or as ``\{.,.\}" according to the
convenience).\par The superJacobian $J({\bf x},{\bf y},{\bf z})$
is
\begin{eqnarray}
\relax J({\bf x},{\bf y},{\bf z}) &=& (-1)^{\epsilon_{\bf
x}\epsilon_{\bf z}}[{\bf x},[{\bf y},{\bf z}]]
+(-1)^{\epsilon_{\bf y}\epsilon_{\bf x}}[{\bf y},[{\bf z},{\bf
x}]]+ (-1)^{\epsilon_{\bf z}\epsilon_{\bf y}}[{\bf z},[{\bf
x},{\bf y}]]. \label{thirteen}
\end{eqnarray} With this in mind, formulas (\ref{six}) can be
reinterpreted as conditions for the superMalcev property.\par In
component level, the superaffine ${\widehat{\cal O}}$ algebra
reads as follows
\begin{eqnarray}
\{\psi_a(x),\psi_b(y)\}&=& k \delta_{ab} \delta(x-y),\nonumber\\
\{ \psi_a (x), j_b(y) \} &=& f_{abc}\psi_c (y)
\delta(x-y),\nonumber\\ \{j_a (x), j_b(y) \} &=&
k\delta_{ab}\partial_y\delta(x-y)+ f_{abc}j_c(y)\delta (x-y).
\label{fourteen}
\end{eqnarray}
Here $j_a(x)$ ad $\psi_a(x)$ are real fields. It should be noticed
that, for later convenience, in the above formulas the $\Psi_0(X)$
superfield has been associated with the $i\tau_0$ octonion. The
real-valued $k$ is the central charge of the superaffine
${\widehat{\cal O}}$ algebra.
\par The proof that (\ref{fourteen}) satisfies the superMalcev
conditions (\ref{six}) consists in a straightforward check.

\vspace{0.2cm} \noindent{\section{The Sugawara construction and
the $N=8$ Non -- associative S.C.A.}}

In this Section we investigate the Sugawara construction performed
with the superaffine fields entering the ${\widehat{\cal O}}$
algebra (\ref{fourteen}). We already know that the purely bosonic
subalgebra admits a Sugawara described in \cite{Osi1}, while it is
possible to prove that the Sugawara associated to the quaternionic
subalgebra (obtained when the octonionic coefficients $a,b$ are
restricted to the values $0,1,2,3$ only) corresponds to the
Sugawara of \cite{IKT} which realizes the $N=4$ ``minimal"
Superconformal Algebra. Both such cases are recovered as a special
limit of the Sugawara construction here described.\par To be
precise we are investigating here the possibility of a closed
algebraic structure in the Enveloping Algebra of ${\widehat{\cal
O}}$, thought as an algebra of {\em real fields} endowed with a
non-Lie super-Poisson bracket structure given by equation
(\ref{fourteen}), assumed to satisfy a graded version of the
Leibniz property, i.e.
\begin{eqnarray}
[{\bf  x} {\bf y},{\bf z}] &=& {\bf x} [{\bf y},{\bf z}]
+(-1)^{\epsilon_{\bf x}\epsilon_{\bf y}}{\bf y} [{\bf x},{\bf z}].
\label{fifteen}
\end{eqnarray}
Specifically, we are looking for an $N=8$ extension of the
Virasoro algebra, covariantly constructed with the imaginary
octonions structure constants (labeled by the greek indices). This
requires a spin-$2$ Virasoro-type field $T$ (scalar w.r.t
imaginary octonions), a fermionic spin-$\frac{3}{2}$ field $Q$
(also scalar w.r.t imaginary octonions) and the extra seven
spin-$\frac{3}{2}$ fields $Q_\alpha$, as well as the seven spin
$1$ bosonic currents $J_\alpha$.\par This problem is well-defined
and admits a complete solution which can be obtained via computer
algebra. We indeed solved it by first rephrasing it in the
language of classical OPE's, which allowed us to perform
computations using Mathematica.\par The final answer is the
following. Unlike both the purely bosonic case and the $N=4$ SCA
quaternionic subalgebra case, no closed algebraic structure can be
found for a given finite value of $k$, the affine central charge
entering (\ref{fourteen}). This is due to the presence of the
extra-terms $(X\star)_{...}$ (\ref{eighteen}), depending on the
$4$-indices structure constant $C_{\alpha\beta\gamma\delta}$ and
the fermionic fields $\psi_\alpha$. All such terms are
automatically vanishing for the two above-mentioned subalgebra
cases.\par However, a closed algebraic structure can be found
after taking a well-definite limit for $k\rightarrow \infty$.\par
Indeed if we set
\begin{eqnarray} T &=& \frac{1}{k^2}(j_aj_a +\psi_a'\psi_a)
+\frac{1}{k}j_0'-\frac{2}{3k^3}
C_{\alpha\beta\gamma}\psi_\alpha\psi_\beta j_\gamma
-\frac{1}{k^4}C_{\alpha\beta\gamma\delta}
\psi_\alpha\psi_\beta\psi_\gamma\psi_\delta,\nonumber\\ Q & =&
\frac{1}{k^2} \psi_aj_a +\frac{1}{k}\psi_0'-\frac{2}{k^3}
C_{\alpha\beta\gamma}\psi_\alpha\psi_\beta\psi_\gamma,\nonumber\\
Q_\alpha &=& \frac{1}{k^2}( \psi_0j_\alpha -\psi_\alpha j_0
-C_{\alpha\beta\gamma}\psi_\beta j_\gamma)
-\frac{1}{k}\psi_\alpha' -\frac{2}{k^3}C_{\alpha\beta\gamma\delta}
\psi_\beta\psi_\gamma\psi_\delta,\nonumber\\ J_\alpha &=&
\frac{1}{k^2} \psi_0\psi_\alpha +\frac{1}{k}j_\alpha -\frac{1}{2
k} C_{\alpha\beta\gamma}\psi_\beta\psi_\gamma \label{sixteen}
\end{eqnarray}
and simultaneously renormalize the Poisson brackets in
(\ref{fourteen}) through
\begin{eqnarray}
\{.,.\} &\mapsto \{.,.\}_R = \frac{k}{2}\{.,.\}, \label{seventeen}
\end{eqnarray}
we obtain a closed algebraic structure, since all the extra-terms
$(X\star)_{...}$ which appear below are vanishing in this
limit.\par The result is in fact
\begin{eqnarray}
\{ T(x), T(y)\}_R &=& -\frac{1}{2} {\partial_y}^3 \delta(x-y) +
2T(y)\partial_y\delta(x-y) +T'(y) \delta(x-y),\nonumber\\ \{T(x),
Q(y)\}_R &=& \frac{3}{2}Q(y)\partial_y\delta (x-y) + Q'(y) \delta
(x-y)+ (X1),\nonumber\\ \{T(x), Q_\alpha(y)\}_R &=&
\frac{3}{2}Q_{\alpha}(y)\partial_y\delta (x-y) + {Q_\alpha}'(y)
\delta (x-y)+ (X2)_\alpha,\nonumber\\
 \{T(x), J_\alpha(y)\}_R &=&
J_{\alpha}(y)\partial_y\delta (x-y) + {J_\alpha}'(y) \delta (x-y)+
(X3)_\alpha,\nonumber\\
 \{Q(x), Q(y)\}_R &=&-\frac{1}{2}{\partial_y}^2\delta(x-y)+
+\frac{1}{2} {T}(y) \delta (x-y)+ (X4),\nonumber\\
 \{Q(x), Q_\alpha(y)\}_R &=&
-J_{\alpha}(y)\partial_y\delta (x-y) -\frac{1}{2} {J_\alpha}'(y)
\delta (x-y)+ (X5)_\alpha,\nonumber\\
 \{Q(x), J_\alpha(y)\}_R &=&
-\frac{1}{2}Q_{\alpha}(y)\delta (x-y)+ (X6)_\alpha,\nonumber\\
 \{Q_\alpha(x), Q_\beta(y)\}_R
 &=&-\frac{1}{2}\delta_{\alpha\beta}{\partial_y}^2\delta(x-y) +
 C_{\alpha\beta\gamma}
 J_\gamma(y)\partial_y\delta(x-y)+\nonumber\\&& +
\frac{1}{2}(\delta_{\alpha\beta}T(y)+C_{\alpha\beta\gamma}
{J_\gamma}'(y))\delta(x-y)+ (X7)_{\alpha\beta},\nonumber\\
 \{Q_\alpha(x), J_\beta(y)\}_R &=&
\frac{1}{2}(\delta_{\alpha\beta} Q(y)-C_{\alpha\beta\gamma}
Q_\gamma (y)) \delta(x-y)+ (X8)_{\alpha\beta},\nonumber\\
 \{J_\alpha(x), J_\beta(y)\}_R &=&
\frac{1}{2}\delta_{\alpha\beta}\partial_y\delta (x-y) -
C_{\alpha\beta\gamma} J_\gamma(y)\delta (x-y)+ (X9)_{\alpha\beta},
\label{eighteen}
\end{eqnarray}
where the extra-terms $(X\star)_{...}$, vanishing in the
$k\rightarrow \infty $ limit, are explicitly given by $
(X\star)_{...} \equiv (\widetilde{X\star})_{...}(y)\delta(x-y)$,
with
\begin{eqnarray}
(\widetilde{X1}) &=& -\frac{3}{2k^4}
C_{\alpha\beta\gamma\delta}\psi_\alpha\psi_\beta\psi_\gamma
j_\delta ,\nonumber\\ (\widetilde{X2})_\alpha &=& \frac{12}{k^5}
C_{\beta\gamma\delta}\psi_0\psi_\alpha\psi_\beta\psi_\gamma\psi_\delta
-\frac{2}{k^4}C_{\alpha\beta\gamma\delta}
\psi_0\psi_\beta\psi_\gamma j_\delta+ \frac{12}{k^5}
C_{\beta\gamma\delta\epsilon}
\psi_\alpha\psi_\beta\psi_\gamma\psi_\delta\psi_\epsilon
+ \nonumber\\ &&+ \frac{4}{k^4} C_{\beta\gamma\delta}
\psi_\alpha\psi_\beta \psi_\gamma j_\delta - \frac{6}{k^4}
\delta_{\alpha\beta} j_\beta C_{\gamma\delta\epsilon}
\psi_\gamma\psi_\delta\psi_\epsilon-\frac{6}{k^4}
C_{\alpha\beta\gamma} \psi_\beta\psi_\gamma j_\delta\psi_\delta
,\nonumber\\ (\widetilde{X3})_\alpha &=& \frac{12}{k^4}
C_{\beta\gamma\delta}\psi_\alpha\psi_\beta\psi_\gamma\psi_\delta
-\frac{2}{k^3} C_{\alpha\beta\gamma\delta} \psi_\beta\psi_\gamma
j_\delta ,\nonumber\\ (\widetilde{X4}) &=& \frac{5}{2k^4}
C_{\alpha\beta\gamma\delta}\psi_\alpha\psi_\beta\psi_\gamma\psi_\delta
,\nonumber\\ (\widetilde{X5})_\alpha &=& \frac{2}{k^4}
C_{\alpha\beta\gamma\delta}\psi_0\psi_\beta\psi_\gamma\psi_\delta
-\frac{6}{k^4} C_{\beta\gamma\delta} \psi_\alpha
\psi_\beta\psi_\gamma\psi_\delta -\frac{1}{3 k^3}
C_{\alpha\beta\gamma\delta} \psi_\beta\psi_\gamma
j_\delta,\nonumber\\ (\widetilde{X6})_\alpha &=& -\frac{7}{3k^3}
C_{\alpha\beta\gamma\delta}\psi_\beta\psi_\gamma\psi_\delta
,\nonumber\\ (\widetilde{X7})_{\alpha\beta} &=& -\frac{4}{k^4}
\delta_{\alpha\beta}
C_{\gamma\delta\epsilon}\psi_0\psi_\gamma\psi_\delta +
\frac{6}{k^4}(C_{\alpha\gamma\delta}
\psi_0\psi_\beta\psi_\gamma\psi_\delta+
C_{\beta\gamma\delta}\psi_0\psi_\alpha\psi_\gamma\psi_\delta)+
\frac{2}{3}C_{\gamma\delta\epsilon} \psi_\gamma\psi_\delta
j_\epsilon -\nonumber\\ && -
\frac{1}{k^3}(C_{\alpha\gamma\delta}\psi_\beta\psi_\gamma
j_\delta+ C_{\beta\gamma\delta} \psi_\alpha\psi_\gamma j_\delta)
-\frac{1}{k^3}(C_{\alpha\gamma\delta} \psi_\gamma\psi_\delta
j_\beta + C_{\beta\gamma\delta}\psi_\gamma\psi_\delta j_\alpha )
+\nonumber \\
 && +\frac{11}{6 k^4}\delta_{\alpha\beta}
C_{\gamma\delta\epsilon\zeta}
\psi_\gamma\psi_\delta\psi_\epsilon\psi_\zeta -\frac{14}{3 k^4}(
C_{\alpha\gamma\delta\epsilon}\psi_\beta\psi_\gamma\psi_\delta
\psi_\epsilon + C_{\beta\gamma\delta\epsilon}
\psi_\alpha\psi_\gamma\psi_\delta\psi_\epsilon ) ,\nonumber\\
(\widetilde{X8})_{\alpha\beta} &=& \frac{7}{3
k^3}\delta_{\alpha\beta}C_{\gamma\delta\epsilon}
\psi_\gamma\psi_\delta\psi_\epsilon-2C_{\beta\gamma\delta}
\psi_\alpha\psi_\gamma\psi_\delta-5C_{\alpha\gamma\delta}\psi_\beta
\psi_\gamma\psi_\delta +
C_{\alpha\beta\gamma\delta}\psi_0\psi_\gamma\psi_\delta +
C_{\alpha\beta\gamma\delta} \psi_\gamma j_\delta, \nonumber\\
(\widetilde{X9})_{\alpha\beta} &=& \frac{2}{k^2}
C_{\alpha\beta\gamma\delta}\psi_\gamma\psi_\delta .
\label{nineteen}
\end{eqnarray}
Some comments are in order. The $k\rightarrow \infty$ limit is
well-defined since the fields in (\ref{sixteen}) are of order
$O(\varepsilon)$ in $\varepsilon = \frac{1}{k}$, while the extra
terms $(X\star)_{...}$ in the r.h.s. of (\ref{eighteen}) are of
higher order in $\varepsilon$. They are indeed of the order
$O(\varepsilon^2)$ or higher. We could have normalized the fields
(\ref{sixteen}) to be of order $O(1)$ in $\varepsilon$, but in
this case the value of the central charge $c$ in the
superconformal algebra would have been $\infty$. The only
possibility of recovering a finite value for the superconformal
central charge $c$ consists in performing the ``classical
renormalization" described above. The value of the conformal
central charge $c$ is unrelated with $k$, the value of the affine
central charge. The formula (\ref{eighteen}) presents for $c$ the
value $c=-6$ ($c$ is obtained as the Virasoro central charge from
the first equation in (\ref{eighteen}) and corresponds to twelve
times the coefficient of $\delta'''$). However, this value can be
normalized at will (since we are dealing with classical Poisson
brackets which satisfy by construction a graded version of the
Leibniz property) through a simultaneous {\em finite} rescaling of
both the fields (collectively denoted as $\phi_i$) in
(\ref{fourteen}) $(\phi_i\mapsto z\phi_i)$ and the Poisson
brackets (\ref{fourteen}) ($\{.,.\} \mapsto \frac{1}{z}\{.,.\}$).
It turns out that $c\mapsto z c$ and in particular we can set
$c=1$ whenever is $c\neq 0$.\par The closed superconformal
algebra, recovered for $(X\star)_{...}\equiv 0$, coincides with
the so-called ``Non-Associative Superconformal $N=8$ Algebra"
introduced for the first time in reference \cite{ESTvPS}.\par The
term ``non-associative" is in reference with the fact that it does
not satisfy a(super)-Jacobi property. It is not even a
(super)Malcev algebra. This point can be understood by noticing
that, under the composition law ${\bf x}= {\bf x}_1\cdot {\bf
x}_2$, the (super)Jacobi property is guaranteed for the triple
${\bf x}, {\bf y}, {\bf z}$, whenever (super)Jacobi is separately
satisfied for the triples ${\bf x}_1, {\bf y}, {\bf z}$ and ${\bf
x}_2, {\bf y}, {\bf z}$ ( $ J({\bf x}_1,{\bf y},{\bf z})=J({\bf
x}_2,{\bf y},{\bf z})= 0 \Rightarrow J({\bf x},{\bf y},{\bf
z})=0)$. On the contrary, the (super)Malcev condition (\ref{six})
is not automatically closed under such a composition law.
Therefore the super-Malcev property satisfied by the superaffine
algebra ${\widehat{\cal O}}$ does not guarantee a super(Malcev)
property for the superconformal algebra extracted through Sugawara
construction. Indeed, an explicit counterexample can be given such
that (\ref{six}) is not verified. It is sufficient to take ${\bf
x} \equiv J_1 (x)$, ${\bf y} \equiv Q_4(y)$, ${\bf z} \equiv
J_2(z)$.
\par On the other hand the bosonic subalgebra,
which is restricted to the fields $T(x)$ and $J_\alpha (x)$ alone,
is a Malcev algebra. \par The existence of the (\ref{sixteen})
Sugawara ensures that the superaffine algebra ${\widehat{\cal O}}$
is compatible with the global $N=8$ supersymmetry. This result is
not surprising and would have been expected since the algebra is
obtained in terms of the octonionic structure constants. In any
case the existence of the (\ref{sixteen}) Sugawara allows us to
explicitly compute the global $N=8$ transformations carried by the
$j_a(x)$ and $\psi_a(x)$ fields entering (\ref{fourteen}). Indeed,
let us introduce the $N=8$ global supersymmetric charges ${\bf
Q}_a$ through \begin{eqnarray} {\bf Q}_a  &=& \oint dx Q_a(x),
\label{twenty}
\end{eqnarray}
where $a=0,1,2,...,7$ and $Q_0(x) \equiv Q(x)$. The supersymmetric
transformation properties for the fields $j_a(x)$, $\psi_a (x)$
are recovered from (\ref{fourteen}) and (\ref{sixteen}) after the
$k\rightarrow \infty $ limit is taken into account. We have
explicitly
\begin{eqnarray}
\delta_{\epsilon_0 {\bf Q}_0} \psi_0 &=& \epsilon_0
\frac{j_0}{2},\nonumber\\ \delta_{\epsilon_0 {\bf Q}_0} j_0 &=&
\epsilon_0 \frac{\psi_0'}{2},\nonumber\\ \delta_{\epsilon_0 {\bf
Q}_0} \psi_\alpha &=& \epsilon_0 \frac{j_\alpha}{2},\nonumber\\
\delta_{\epsilon_0 {\bf Q}_0} j_\alpha &=& \epsilon_0 (
\frac{\psi_\alpha '}{2}-\frac{2}{k^2} C_{\alpha\beta\gamma\delta}
\psi_\beta\psi_\gamma\psi_\delta )
 ,\nonumber\\
\delta_{\epsilon_\alpha {\bf Q}_\alpha} \psi_0 &=& \epsilon_\alpha
\frac{j_\alpha}{2},\nonumber\\ \delta_{\epsilon_\alpha {\bf
Q}_\alpha} j_0 &=& -\epsilon_\alpha \frac{\psi_\alpha '
}{2},\nonumber\\ \delta_{\epsilon_\alpha {\bf Q}_\alpha}
\psi_\beta &=& \epsilon_\alpha (-\delta_{\alpha\beta}
\frac{j_0}{2} -\frac{1}{2} C_{\alpha\beta\gamma} j_\gamma
-\frac{1}{k} (\psi_\alpha\psi_\beta +
C_{\alpha\beta\gamma}\psi_0\psi_\gamma )) ,\nonumber\\
\delta_{\epsilon_\alpha {\bf Q}_\alpha} j_\beta &=&
\epsilon_\alpha (\delta_{\alpha\beta} \frac{\psi_0 '}{2} +
\frac{1}{2} C_{\alpha\beta\gamma} \psi_\gamma ' + \frac{1}{k}
(C_{\alpha\beta\gamma} \psi_\gamma j_0 -C_{\alpha\beta\gamma}
\psi_0\psi_\gamma +\psi_\alpha j_\beta -\psi_\beta j_\alpha
+C_{\alpha\beta\gamma\delta} \psi_\gamma\psi_\delta) +\nonumber\\
&& +\frac{2}{k^2} (\delta_{\alpha\beta} C_{\gamma\delta\epsilon}
\psi_\gamma\psi_\delta\psi_\epsilon -C_{\beta\gamma\delta}
\psi_\alpha\gamma\delta -2C_{\alpha\gamma\delta}
\psi_\beta\psi_\gamma\psi_\delta )).
\end{eqnarray}
The global supersymmetric transformations, dependent on the
infinitesimal fermionic parameter $\epsilon_a$, are recovered from
the above formulas after taking the $k\rightarrow \infty$ limit.
Please notice that no summation is made for what concerns the
index $\alpha$.

\vspace{0.2cm}\noindent{\section{Conclusions}}

In this paper we have presented the generalization to the
supersymmetric case of some results due to Osipov \cite{Osi1}
concerning the bosonic Malcev affinization of the commutator
algebra of the octonions and its Sugawara construction. We have
introduced the superaffine Malcev algebra ${\widehat{\cal O}}$,
realized by eight bosonic spin-$1$ and eight fermionic
spin-$\frac{1}{2}$ fields, and have been able to prove that a
Sugawara construction exists which maps, after a suitable limit is
taken, these fields into the $N=8$ non-associative superconformal
algebra of reference \cite{ESTvPS}.\par As a corollary, the
superaffine algebra ${\widehat{\cal O}}$, besides being manifestly
$N=1$ supersymmetric, is globally $N=8$ supersymmetric.\par Such
an algebraic construction can find a variety of applications (some
of them are currently under investigation) to a full class of
physical problems involving $N=8$ extended supersymmetries.\par
E.g. the superaffine ${\widehat{\cal O}}$ algebra defines a
Poisson bracket structure underlining the $N=8$ supersymmetric
extension of the $NLS-mKdV$ equations, while the Sugawara-induced
non-associative superconformal algebra gives the Poisson brackets
for an $N=8$ extension of KdV. This is the content of a
forthcoming paper \cite{CRT2} which extends the construction of
\cite{IKT} to the present case.\par Further topics of
investigation concern the geometric approach (see \cite{BSV}) to
the classical superstring dynamics. It is expected that the motion
of a classical superstring in a $10$-dimensional target could be
reduced to a octonionic valued superLiouville theory, which should
be naturally described through the superalgebras introduced here.
On the other hand, either WZNW-type models defined in the ``almost
group manifold $S^7$" (the almost being referred to the fact that
it can be recovered from imaginary octonions, whose product
however is non-associative) and the dynamics of generalized tops
moving on $S^7$ are likely to be described by (super)-Malcev
algebras. A converse approach, which retains Jacobi identities at
the ´price of making field-dependent the algebraic structure
constants (soft-algebras) has been suggested in \cite{CePr}.
Another natural field of investigation concerns the twistor
formalism applied to the Green-Schwarz string (see e.g.
\cite{Ber}).

\end{document}